\documentclass[12pt]{article}
\usepackage{amsfonts,a4}
\usepackage{epsfig}
\usepackage{diss}

\newcommand{\F}{\mathcal F}
\newcommand{\One}{\mathbb I}
\newcommand{\lie}[1]{\mathop{\rm #1}\nolimits}
\newcommand{\spann}{\mathop{\rm span}\nolimits}
\begin{document}

\title{On the equivalence of the discrete nonlinear Schr\"odinger
  equation and the discrete isotropic Heisenberg magnet.}
\author{Tim Hoffmann\\
  \small Fachbereich Mathematik MA 8-3,Technische Universit\"at Berlin\\[-5pt]
  \small Stra\ss e des 17. Juni 136\\[-5pt]
  \small10623 Berlin\\[-5pt]
  \small email: timh@sfb288.math.tu-berlin.de\\[-5pt]
  \small phone: +49-30-314 25784\\[-5pt]
  \small fax: +49 - 30 - 314 21577 } \maketitle
\begin{abstract}
  The equivalence of the discrete isotropic Heisenberg magnet (IHM)
  model and the discrete nonlinear Schr\"odinger equation (NLSE) given
  by Ablowitz and Ladik is shown. This is used to derive the
  equivalence of their discretization with the one by Izgerin and
  Korepin. Moreover a doubly discrete IHM is presented that is
  equivalent to Ablowitz' and Ladiks doubly discrete NLSE.
\end{abstract}
\section{Introduction}
\label{gauge:sec:introduction}
The gauge equivalence of the continuous isotropic Heisenberg magnet
model and the nonlinear Schr\"odinger equation is well known
\cite{FT86}. On the other hand there are several discretizations of
the nonlinear Schr\"odinger equation in literature
(e.\ g.\ . \cite{AL76a,IK81,DJM82d,QNCL84}). In particular there are two famous
versions with continuous time. One introduced by Ablowitz and Ladik
\cite{AL76a} (from now on called dNLSE$\hbox{}_{AL}$) and one given by Izgerin and
Korepin \cite{IK81} (from now on referred to as dNLSE$\hbox{}_{IK}$) (see also
\cite{FT86}). The second can be obtained from the discrete (or lattice)
isotropic Heisenberg magnet model (dIHM) with slight modification via
a gauge transformation \cite{FT86}.

In this paper the gauge equivalence of the dIHM model and the dNLSE$\hbox{}_{AL}$
is shown. In fact this is in complete analogy to the continuous case.
The equivalence of the two discretizations of the nonlinear
Schr\"odinger equation is derived from this.

In addition in Section~\ref{gauge:sec:ddIHM} a doubly discrete (with
discrete time) version of the IHM model is given that links in the
same way with the doubly discrete NLSE introduced by Ablowitz and
Ladik in \cite{AL76b}. It first
appeared in a somewhat implicit form in \cite{DJM82a,QNCL84}.

In \cite{Ho99} the author explains the geometric background of the
interplay between IHM model and NLSE (see also \cite{BS98,DS99}) From
the geometric point of view the dNLSE$\hbox{}_{AL}$ seems to be the
more natural choice.

In the following we will identify $\R^3$ with $\lie{su}(2)$ that is the
span of $\I,\J,$ and $\K$ where 
\[ \I = i\sigma_3 = \quadmatrix{i}{0}{0}{-i}\quad \J = i\sigma_1 =
\quadmatrix{0}{i}{i}{0}\]
\[\K = -i\sigma_2 = \quadmatrix{0}{-1}{1}{0}\]


\section{Equivalence of the discrete Heisenberg magnetic model and the nonlinear Schr\"odinger equation}
\label{gauge:sec:discreteTheory}
The dIHM model and the dNLSE$\hbox{}_{AL}$ are well known
\cite{AL76a,FT86,BS98,SU97}. In this section it is shown that---as in the
smooth case---both models are gauge equivalent.
We start by giving the discretizations.

The dNLSE$\hbox{}_{AL}$ has the form:
\begin{equation}
  \label{gauge:eq:dNLSE}
  -i \dot\Psi_k = \Psi_{k+1} - 2 \Psi_k + \Psi_{k-1} +\vert\Psi_k\vert^2
  (\Psi_{k+1} + \Psi_{k-1})
\end{equation}
It has the following zero curvature representation (see \cite{AL76a,SU97})
\begin{equation}
  \label{gauge:eq:zerodNLSE}
  \dot {\hat L}_k = \hat M_{k+1} \hat L_k - \hat L_k \hat M_k
\end{equation}
with $\hat L_k$ and $\hat M_k$ of the form 
\begin{equation}
  \label{gauge:eq:dNLSELaxPair}
  \begin{array}{rcl}
    \hat L_k(\mu) &=& \quadmatrix{\mu}{\Psi_k}{-\bar\Psi_k}{\mu^{-1}}\\[0.6cm]
    \hat M_k(\mu) &=& \quadmatrix{\mu^2i - i + i\Psi_k\bar\Psi_{k-1}}{\mu i\Psi_k
    - \mu^{-1} i\Psi_{k-1}}{-\mu i\bar\Psi_{k-1} +
    \mu^{-1}i\bar\Psi_k}{-\mu^{-2}i + i -i\bar\Psi_k\Psi_{k-1}}
  \end{array}
\end{equation}
were $\bar{\phantom{x}}$ denotes complex conjugation.
Aiming to the forthcoming theorem we gauge this Lax pair with
$\quadmatrix{\sqrt{\mu}}{0}{0}{\sqrt{\mu}^{-1}}$ and get 
\begin{equation}
  \label{gauge:eq:dNLSELaxPair2}
  \begin{array}{rcl}
    L_k(\mu) &=&
    \quadmatrix{1}{\Psi_k}{-\bar\Psi_k}{1}\quadmatrix{\mu}{0}{0}{\mu^{-1}}
    \\[0.6cm]
    M_k(\mu) &=&
    \quadmatrix{i\Psi_k\bar\Psi_{k-1}}{i\Psi_k -
    i\Psi_{k-1}}{-i\bar\Psi_{k-1} +
    i\bar\Psi_k}{-i\bar\Psi_k\Psi_{k-1}} +\\[0.6cm]
    &&+
    \quadmatrix{1}{\Psi_{k-1}}{-\bar\Psi_{k-1}}{1}\quadmatrix{i(\mu^2 
    - 1)}{0}{0}{-i(\mu^{-2} -1)}.
  \end{array}
\end{equation}

We now turn our attention for a moment to the discrete isotropic
Heisenberg magnet model. It is given by the following evolution
equation
\begin{equation}
  \label{gauge:eq:dIHM}
  \dot S_k = 2\frac{S_{k+1}\times S_k}{1 +\<S_{k+1},S_k>} -
  2\frac{S_{k}\times S_{k-1}}{1 +\<S_{k},S_{k-1}>}
\end{equation}
with the $S_k$ being unit vectors in $\R^3$. Its zero curvature
representation is given by 
\begin{equation}
  \label{gauge:eq:zerodIHM}
  \dot U_k = V_{k+1}U_k - U_kV_k
\end{equation}
with $U_k$ and $V_k$ of the form 
\begin{equation}
  \label{gauge:eq:dIHMLaxPair}
  \begin{array}{rcl}
    U_k &=& \One + \lambda S_k\\[0.2cm]
    V_k &=& -\frac1{1+\lambda^2}\left(2 \lambda^2 \frac{S_{k} +
    S_{k-1}}{1 + \<S_{k},S_{k-1}>} + 2 \lambda \frac{S_{k}\times
    S_{k-1}}{1 +\<S_{k},S_{k-1}>} \right)
  \end{array}
\end{equation}
if one identifies the $\R^3$ with $\lie{su}(2)$ in the usual way.
Now we are prepared to state
\begin{theorem}\label{gauge:thm:dIHMdNLSEequivalence}
  The discrete nonlinear Schr\"odinger equation dNLSE$\hbox{}_{AL}$
  (\ref{gauge:eq:dNLSE}) and the discrete isotropic Heisenberg magnet
  model dIHM (\ref{gauge:eq:dIHM}) are gauge equivalent.
\end{theorem}
\begin{proof}
  We use the notation introduced above. Let $\F$ be a solution to the
  linear problem
  \begin{equation}
    \label{gauge:eq:dNLSElinearProblem}
    \F_{k+1} = L_k(1)\F_k, \quad \dot\F_k = \hat M_k(1)\F_k :=
      \left(M_k(1) + \F_kc\F_k^{-1}\right)\F_k 
  \end{equation}
  with a constant vector $c$. Since $\hat M_{k+1}(1)L_k(1) -
  L_k(1)\hat M_k(1) = M_{k+1}(1)L_k(1) - L_k(1)M_k(1) = \dot L_k(1)$
  the zero curvature condition stays valid and the system is solvable.
  The additional term $ \F_kc\F_k^{-1}$ will give rise to an
  additional rotation around $c$ in the dIHM model. The importance
  of this possibility will be clarified in the next section.  Moreover
  define
  \begin{equation}
    \label{gauge:eq:SfromPsi}
    S_k := \F_k^{-1}\I\F_k.  
  \end{equation}
  Note that this implies that 
  \begin{equation}
    \label{gauge:eq:normPsi}
    \frac{\vert S_k\times S_{k+1}\vert}{1 + \<S_k,S_{k+1}>} = \vert\Psi_k\vert.
  \end{equation} In other words: $\vert\Psi_k\vert = \tan(\frac{\phi_k}2)$
  with $\phi_k = \angle(S_k, S_{k+1})$. We will show, that the $S_k$ solve
  the dIHM model (if $c = 0$). To do so we use $\F^{-1}$ as a gauge field:
  \[ L^{\F^{-1}}_k(\mu) := \F^{-1}_{k+1}L_k(\mu)\F_k =
  \F^{-1}_{k}\quadmatrix{\mu}{0}{0}{\mu^{-1}}\F_k\]
  If one writes $\mu = \sqrt\frac{1 + i\lambda}{1 - i\lambda} = \frac{1
    + i \lambda}{\sqrt{1 + \lambda^2}}$ one gets $\mu^{-1} = \frac{1
    - i \lambda}{\sqrt{1 + \lambda^2}}$ and one can conclude that
  \begin{equation}
    \label{gauge:eq:Lgauged}
    L^{\F^{-1}}_k(\lambda) = \F^{-1}_{k} \frac{\One + \I\lambda}{\sqrt{1 +
        \lambda^2}}\F_k = \frac1{\sqrt{1 + \lambda^2}}(\One + \lambda S_k)
  \end{equation}
  This clearly coincides with $U_k(\lambda)$ up to the irrelevant
  normalization factor $\frac1{\sqrt{1 + \lambda^2}}$.
  On the other hand one gets for the gauge transform of $M_k(\mu)$
  \[ 
  \begin{array}{rcl}
    M^{\F^{-1}}_k(\mu) &:=& \F^{-1}_kM_k(\mu)\F_k - \F^{-1}_k\dot\F_k = 
    \F^{-1}_k\left(M_k(\mu) - M_k(1)- \F_kc\F_k^{-1}\right)\F_k\\[0.5cm]
    &=& \F^{-1}_kL_{k-1}(1)\F_k\F^{-1}_k\quadmatrix{i(\mu^2 
      - 1)}{0}{0}{-i(\mu^{-2} -1)}\F_k - c
  \end{array}\]
  But with above substitution for $\mu$ one gets 
  \begin{equation}
    \label{gauge:eq:mu}
    \quadmatrix{i(\mu^2 - 1)}{0}{0}{-i(\mu^{-2} -1)} = -2
    \frac{\lambda\One + \lambda^2\I}{1+\lambda^2}
  \end{equation}
  and since $\F_k^{-1}L_{k-1}(1)\F_k = \F_{k-1}^{-1} L_{k-1}(1)\F_{k-1}$ we
  get 
  \[
  \begin{array}{rl}
    \F_k^{-1}L_{k-1}(1)\F_k &= \One +
    \F_{k-1}^{-1}\left(\Im(\Psi_{k-1})\J-\Re(\Psi_{k-1})\K\right)\F_{k-1}\\
    &= \One + \F_{k}^{-1}\left(\Im(\Psi_{k-1})\J-\Re(\Psi_{k-1})\K\right)\F_{k}
  \end{array}
  \]
  Remember that $S_k = \F_{k}^{-1}\I\F_{k}$ and $S_{k-1} =
  \F_{k-1}^{-1}\I\F_{k-1}$. Using equation (\ref{gauge:eq:normPsi}) and
  the fact that $\I$ and $\Im(\Psi_{k-1})\J-\Re(\Psi_{k-1})\K$
  anti-commute we conclude\footnote{to fix the sign of the second term
    one needs to look at   the sign of the scalar product
    $\<\F_{k}^{-1}\left(\Im(\Psi_{k-1})\J -
      \Re(\Psi_{k-1})\K\right)\F_{k}, \frac{S_{k} \times S_{k-1}}{1 +
      \<S_{k},S_{k-1}>}>$.}  
  \begin{equation}
    \label{gauge:eq:cross}
    \F_k^{-1}L_{k-1}(1)\F_k = \One + \frac{S_{k}\times S_{k-1}}{1 +
      \<S_{k},S_{k-1}>} 
  \end{equation}
  Combining this and equation (\ref{gauge:eq:mu}) one obtains for the
  gauge transform of $M_k$
  \begin{equation}
    \label{gauge:eq:gaugeM}
    \begin{array}{rl}
      M^{\F^{-1}}_k(\lambda) =& -2\left( \One + \frac{S_{k}\times S_{k-1}}{1 +
          \<S_{k},S_{k-1}>}\right) \frac{\lambda\One +
        \lambda^2S_k}{1+\lambda^2} -c \\[0.4cm]
      &= \frac{-2}{1+\lambda^2}\left(\lambda\One +
        \lambda\frac{S_{k}\times S_{k-1}}{1 + \<S_{k},S_{k-1}>}
        +\lambda^2\left(S_k + \frac{(S_{k}\times S_{k-1})S_k}{1 +
            \<S_{k},S_{k-1}>}\right)
      \right) - c \\[0.4cm]
      &= \frac{-2\lambda}{1+\lambda^2}\One - \frac{2}{1+\lambda^2}\left(
        \lambda\frac{S_{k}\times S_{k-1}}{1 + \<S_{k},S_{k-1}>}
        +\lambda^2\frac{S_k + S_{k-1}}{1 + \<S_{k},S_{k-1}>}\right) -
          c \\[0.4cm]
      &= \frac{-2\lambda}{1+\lambda^2}\One + V_k(\lambda) - c
    \end{array}
  \end{equation}
  Since the first term is a multiple of the identity and independent of
  $k$ it cancels in the zero curvature condition and therefore can be
  dropped. This gives the desired result if $c = 0$.
\end{proof}

\subsection{Equivalence of the two discrete nonlinear Schr\"odinger
  equations}
\label{gauge:sec:LNS1vsLNS2}

There has been another discretization of the nonlinear Schr\"odinger
equation in the literature \cite{IK81,FT86}. It can be derived from a
slightly modified dIHM model by a gauge transformation. Since we
showed that the dNLSE$\hbox{}_{AL}$ introduced by Ablowitz and Ladik is gauge
equivalent to the dIHM it is a corollary of the last theorem that the
two discretizations of the NLSE are in fact equivalent.

The method of getting the variables of this other discretization is
basically a stereographic projection of the variables $S_k$ from the
dIHM \cite{FT86}: One defines
\begin{equation}
  \label{gauge:eq:chis}
  \chi_k = \chi(S_k) = \sqrt{2}(-1)^k\frac{2(S_k +\I) -\vert S_k
  +\I\vert^2\I}{\sqrt{\vert S_k  +\I\vert^4 + \vert 2(S_k +\I) -\vert
  S_k +\I\vert^2\I \vert^2}}
\end{equation}
or 
\begin{equation}
  \label{gauge:eq:chis2}
  S_k = (1 - \vert\chi_k\vert^2)\I + \Im\left(\sqrt{2}(-1)^k\chi_k \sqrt{1
    - \frac{\vert\chi_k\vert^2}2}\right) \J -\Re\left(\sqrt{2}(-1)^k\chi_k \sqrt{1
    - \frac{\vert\chi_k\vert^2}2}\right) \K
\end{equation}
If one modifies the evolution (\ref{gauge:eq:dIHM}) by adding a rotation
around $\I$
\begin{equation}
  \label{gauge:eq:modifieddIHM}
  \dot S_k = 2\frac{S_{k+1}\times S_k}{1 +\<S_{k+1},S_k>} -
  2\frac{S_{k}\times S_{k-1}}{1 +\<S_{k},S_{k-1}>} - 4 S_k\times \I
\end{equation}
writing this in terms of the new variables $\chi_k$ gives rise to the
following evolution equation (dNLSE$\hbox{}_{IK}$):
\begin{equation}
  \label{gauge:eq:vardNLSE}
  -i\dot\chi_k = 4\chi_k + \frac{P_{k,k+1}}{Q_{k,k+1}} +
  \frac{P_{k,k-1}}{Q_{k,k-1}} 
\end{equation}
where 
\[
\begin{array}{rl}
P_{n,m} =& -\left(\chi_n + \chi_m \sqrt{1 -\frac{|\chi_n|^2}2}
  \sqrt{1 -\frac{|\chi_m|^2}2} - \chi_n|\chi_m|^2 -\right.\\
  &\left.-\frac14(|\chi_n|^2\chi_m + \chi_n^2\bar \chi_m) \frac{\sqrt{1
  -\frac{|\chi_m|^2}2}}{\sqrt{1 -\frac{|\chi_n|^2}2}}\right)
\end{array}
\] 
and 
\[
\begin{array}{rcl}
  Q_{n,m} & = &1 -\frac12\bigg(|\chi_n|^2 + |\chi_m|^2 +(\chi_n\bar\chi_m
  + \bar\chi_n\chi_m)\sqrt{1 -\frac{|\chi_n|^2}2} \sqrt{1
  -\frac{|\chi_m|^2}2} -\\ & &
  -|\chi_n|^2|\chi_m|^2\bigg).
\end{array}
\]
This evolution clearly possesses a zero curvature condition $\dot U_k
= \hat V_{k+1}U_k - U_k\hat V_k$ with
\begin{equation}
  \label{gauge:eq:dNLSE2LaxPair}
  \hat V_k(\lambda) = V_k(\lambda) - 2\I
\end{equation}
since one can view $S_k$ as a function of
$\chi_{k}$ via equation (\ref{gauge:eq:chis2}).
\begin{theorem}
  The dNLSE$\hbox{}_{IK}$ (\ref{gauge:eq:vardNLSE})and the dNLSE$\hbox{}_{AL}$
  (\ref{gauge:eq:dNLSE}) are gauge equivalent.
\end{theorem}
\begin{proof}
  This is already covered by the proof of
  theorem~\ref{gauge:thm:dIHMdNLSEequivalence}.
\end{proof}
Since the $S_k$ are given by $S_k = \F_k^{-1}\I\F_k$ the $\chi_k$ are
functions of the $\Psi_k$ and vice versa, but these maps are nonlocal.

\section{A doubly discrete IHM model and the doubly discrete NLSE}
\label{gauge:sec:ddIHM}

In the following we will construct a discrete time evolution for the
variables $S_k$ that---applied twice---can be viewed as a doubly discrete
IHM model. In fact it will turn out that this system is equivalent to
the doubly discrete NLSE introduced by Ablowitz and Ladik
\cite{AL76b}. We start by defining the zero curvature representation.
\begin{equation}
  \label{gauge:eq:ddIHMLaxPair}
  \begin{array}{rcl}
  U_k(\lambda) = \One + \lambda S_k\\[0.4cm]
  V_k(\lambda) = \One + \lambda(r\One + v_k)
  \end{array}
\end{equation}
with $r\in\R$. The $v_k$ (as well as the $S_k$) are vectors in $\R^3$
(again written as complex 2 by 2 matrix). The zero curvature condition
$\tilde L_k V_{k} = V_{k+1}L_k$ should hold for all $\lambda$ giving
$v_k + \tilde S_k = S_k + v_{k+1}$ and $r(\tilde S_k - S_k) =
v_{k+1}S_k - \tilde S_k v_k$. (Here and in the forthcoming we use
$\tilde{\phantom{x}}$ to denote the time shift.) One can solve this
for $v_{k+1}$ or $\tilde S_k$ getting
\begin{equation}
  \label{gauge:eq:vEvol}
  \begin{array}{rcl}
  v_{k+1} & = & (S_k - v_k -r) v_k (S_k - v_k -r)^{-1}\\
  \tilde S_k & = & (S_k - v_k -r) S_k (S_k - v_k -r)^{-1}
  \end{array}
  \end{equation}
  This can be interpreted in the following way:
  Since $S_k,v_{k+1},-\tilde S_k,$ and $-v_k$ sum up to zero they can
  be viewed as a quadrilateral in $\R^3$. But equation
  (\ref{gauge:eq:vEvol}) says that $v_{k+1}$ and $\tilde S_k$ are
  rotations\footnote{Any rotation of a vector $v$ in
    $\R^3=\lie{su}(2)$ can be written as conjugation with a matrix
    $\sigma$ of the form $\sigma = \cos(\frac\phi2)\One +
    \sin(\frac\phi2)a$ where $\phi$ is the rotation angle and $a$ the
    rotation axis with $\vert a\vert = 1$.} of $v_k$ and $S_k$ around
  $S_k - v_k$. So the resulting quadrilateral is a parallelogram that
  is folded along one diagonal.  See \cite{Ho99} to get a more
  elaborate investigation of the underlying geometry.
  
  Equation (\ref{gauge:eq:vEvol}) is still a
  transformation\footnote{In fact it is the B\"acklund transformation
    for the dIHM model!}  and no evolution since one has to fix an
  initial $v_0$. But in the case of periodic $S_k$ one can find in
  general two fix points of the transport of $v_0$ once around the
  period and thus single out certain solutions.  If on the other hand
  one has rapidly decreasing boundary conditions one can extract
  solutions by the condition that $\tilde S_k \to \pm S_k$ for
  $k\to\infty$ and $k\to -\infty$. But instead of going into this we
  will show, that doing this transformation twice is equivalent to
  Ablowitz' and Ladiks system.

Let us recall their results.
\begin{theorem}[Ablowitz and Ladik 77]
\label{gauge:thm:AblowitzLadik}
  Given the matrices 
  \[L_k(\mu) = \quadmatrix{1}{\Psi_k}{-\bar \Psi_k}{1}
  \quadmatrix{\mu}{0}{0}{\mu^{-1}}\]
  and $V_k(\mu)$ with the following $\mu$--dependency:
  \[V_k(\mu) = \mu^{-2} V^{(-2)}_k + 
  V^{(0)}_k + 
  \mu^{2} V^{(2)}_k\] with $V^{(-2)}_k$ being upper and $V^{(2)}_k$
  being lower triangular.  Then the zero curvature condition
  $V_{k+1}(\mu)L_k(\mu) = \tilde L_k(\mu) V_k(\mu)$ gives the
  following equations:
\begin{equation}
  \label{sr:alevolution}
  \begin{array}{rcl}
    (\tilde \Psi_k - \Psi_k)/i &=& \alpha_+ \Psi_{k+1} - \alpha_0
    \Psi_k + \bar\alpha_0 \tilde \Psi_k - \bar\alpha_+ \tilde
    \Psi_{k-1} + (\alpha_+ \Psi_k \mathcal A_{k+1} -\\[0.2cm]
    &&-\bar\alpha_+\tilde \Psi_k \bar{\mathcal A}_k) + (-\bar\alpha_-
    \tilde \Psi_{k+1} + \alpha_- \Psi_{k-1})(1 + |\tilde \Psi_k|^2
    )\Lambda_k\\[0.4cm]
    \mathcal A_{k+1} - \mathcal A_k&=& \tilde \Psi_k \bar{\tilde \Psi}_{k-1}
    - \Psi_{k+1}\bar \Psi_k\\[0.2cm]
    \Lambda_{k+1}(1+|\Psi_k|^2) &=& \Lambda_{k}(1+|\tilde \Psi_k|^2)
  \end{array}
\end{equation}
with constants $\alpha_+,$ $\alpha_0$ and $\alpha_-$.

In the case of
periodic or rapidly decreasing boundary conditions the 
natural conditions $ \mathcal A_k \rightarrow 0,$ and
$\Lambda_k\rightarrow 1$ 
for $k\rightarrow \pm\infty$ give formulas for $\mathcal A_k$ and
$\Lambda_k$:
\[ \mathcal A_k = \Psi_k \bar \Psi_{k-1} + \sum_{j = j_0}^{k-1} (\Psi_j\bar
\Psi_{j-1} - \tilde \Psi_j\bar{\tilde \Psi}_{j-1})\]
\[ \Lambda_k = \prod_{j = j_0}^{k-1}\frac{1 + |\tilde
  \Psi_j|^2}{1+|\Psi_j|^2}\] 
with $j_0 = 0$ in the periodic case and $j_0 = -\infty$ in case of
rapidly decreasing boundary conditions.
\end{theorem}
Note that this is not the most general version of their result. One
can make $\Psi$ and $\bar\Psi$ independent variables which results in
slightly more complicated equations but the given reduction to the
NLSE case is sufficient for our purpose.

\begin{theorem}
  The system obtained by applying the above transformation twice is
  equivalent to the doubly discrete Ablowitz Ladik system in
  theorem~\ref{gauge:thm:AblowitzLadik}.
\end{theorem}
\begin{proof}
  The method is more or less the same as in the singly discrete
  case although this time we start from the other side:
  
  Start with a solution $S_k$ of the ddIHM model. Choose $\F_k$ such
  that  
  \begin{equation}
    \label{gauge:eq:discreteParallelFrame}
    \begin{array}{rcl}
    \F_k^{-1}\I\F_k & = & S_k\\[0.4cm]
    \left[(\F_{k+1}^{-1}\J\F_{k+1}),(\F_{k}^{-1}\J\F_{k}) \right] & \|
    & \left[ S_{k+1},S_k  \right]
    \end{array}
  \end{equation}
  This is always possible since the first equation leaves a
  gauge freedom of rotating around $\I$. Moreover define $L_k(1) =
  \F_{k+1}\F_k^{-1}$ and normalize $\F_k$ in such a way that $L_k(1)$
  takes the form
  \[ L_k(1) = \One + A_k\]
  equations~(\ref{gauge:eq:discreteParallelFrame}) ensure  that
  $A_k\in\spann(\J,\K)$ and thus can be written $A_k = \Re(\Psi_k)\K -
  \Im(\Psi_k)\J$ for some complex $\Psi_k$. Equipped with this we can
  gauge a normalized version of $M_k(\lambda)$ with $\F_k$ and get 
  \begin{equation}
    \label{gauge:eq:gaugedDiscreteM}
    \begin{array}{rcl}
      M_k^{\F} & = & \frac1{\sqrt{1+\lambda^2}}
      \F_{k+1}M_k(\lambda)\F_k^{-1} = L_k(1)\frac{\One +
        \lambda\I}{\sqrt{1+\lambda^2}}\\
      & = & \quadmatrix{1}{\Psi_k}{-\bar\Psi_k}{1}
      \quadmatrix{\mu}{0}{0}{\mu^{-1}} 
    \end{array}
  \end{equation}
  if we write $\mu = \frac{1 + i\lambda}{\sqrt{1 + \lambda^2}}$ as
  before. On the other hand we get for an---again
  renormalized---$N_k(\lambda)$  
  \begin{equation}
    \label{gauge:eq:gaugedDiscreteN}
    \begin{array}{rcl}
    N_k^{\F} & = & \frac{1+\mu^2}{\mu}\tilde\F_k
    N_k(\lambda)\F_k^{-1} = (\frac1\mu +\mu)\tilde\F_k\F_k^{-1} +
    (\frac1\mu - \mu)\tilde\F_k(r + v_k)\F_k^{-1} \\
    & = & \mu^{-1} V_k^- + \mu V_k^+
    \end{array}
  \end{equation}
  But the zero curvature condition $\tilde L_k(\mu)N_k^{\F}(\mu) =
  N_{k+1}^{\F_k}(\mu)L_k(\mu)$ yields that $V_k^+$ must be lower
  and $V_k^-$ upper triangular. Thus $\tilde
  N_k^{\F}(\mu)N_k^{\F}(\mu)$ has the $\mu$-dependency as required in
  Theorem~\ref{gauge:thm:AblowitzLadik}.
\end{proof}


\subsection*{Acknowledgments}
The author would like to thank A. Bobenko and Y. Suris for helpful
discussions and encouragement.

\bibliographystyle{plain}
\bibliography{discrete}
\end{document}